\documentclass[10pt,letterpaper,twocolumn]{article} 

\usepackage{ol2}
\usepackage[draft]{hyperref}
\usepackage{amsmath}

\usepackage{graphicx}
\usepackage{color}

\begin{document}

\twocolumn[

\title{A compact high-flux cold atom beam source}

\author{James R. Kellogg,$^{1,*}$ Dennis Schlippert,${^2}$ James M. Kohel,${^1}$ Robert J. Thompson,${^1}$ David C. Aveline,${^1}$ and Nan Yu${^1}$}
\address{
$^1$Jet Propulsion Laboratory, California Institute of Technology, \\ 4800 Oak Grove Drive, Pasadena, CA 91109-8099, USA \\
$^2$Institut f\"{u}r Quantenoptik, Leibniz Universit\"{a}t Hannover, \\ Welfengarten 1, 30167 Hannover, Germany \\
$^*$Corresponding author: james.r.kellogg@jpl.nasa.gov
}

\begin{abstract}We report on an efficient and compact high-flux Cs atom beam source based on a retro-reflected two-dimensional magneto-optical trap (2D~MOT). We realize an effective pushing field component by tilting the 2D~MOT collimators towards a separate three-dimensional magneto-optical trap (3D~MOT) in ultra-high vacuum. This technique significantly improved 3D~MOT loading rates to greater than $8 \times 10^9$ atoms/s using only 20~mW of total laser power for the source. When operating below saturation, we achieve a maximum efficiency of $6.2 \times 10^{11}$ atoms/s/W.
\end{abstract}


]


\begin{figure}
\includegraphics[width=0.95\linewidth]{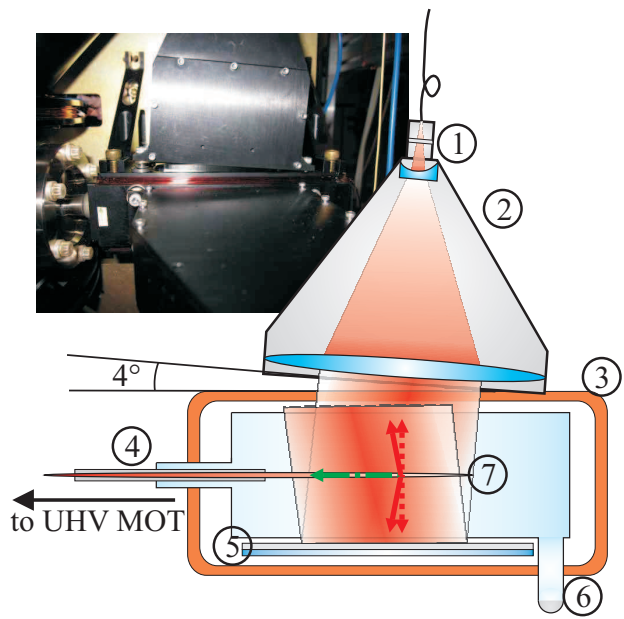}
\caption{Schematic illustration of the experimental setup of our 2D~MOT source  (not to scale) with $\lambda/4$-wave plate (1), collimator (2), orientation of the current coils (3), differential pumping tube (4), retro-reflection optics (5), Cs reservoir inside completely sealed glass cuvette (6), and the atomic beam (7) (the setup for the orthogonal axis is identical and is not shown for clarity). Radiation forces acting upon the atoms are illustrated (red arrows) and split into a transversal cooling component (red dashed arrows) and an effective pushing component (green dash-dotted arrow; not to scale). The collimation lens for the slowly diverging axis is not displayed. The inset shows a photo of 2D~MOT collimators depicting $4^{\circ}$ angle with respect to the axis of zero magnetic field (UHV region at left).}
\label{apparatus}
\end{figure}

The performance of cold atom experiments relying on three-dimensional magneto-optical trap (3D~MOT)~\cite{
RaabPRL87,SteaneJOptSoc92} techniques can be greatly enhanced by employing a high-flux cold atom beam to obtain high atom loading rates while maintaining low background pressures in the UHV MOT regions. Several techniques exist for generating slow beams of cold atoms ~\cite{PhillipsPRL82,LuPRL96,DieckmannPRA98,GhaffariPRA99}. However, one of the technically simplest approaches is a two-dimensional magneto-optical trap (2D~MOT)~\cite{WeyersOptComm97,BerthoudEurPhys98,SchoserPRA02,Ramirez-SerranoOL06,MuellerPRA07}. J\"{o}llenbeck et al.\ recently reported a loading rate of $8.4 \times 10^{10}$ atoms/s realized using 0.6~W of laser power in a 2D$^+$-MOT~\cite{JoellenbeckPRA11}, yielding an efficiency of $2.7 \times 10^{11}$ atoms/s/W. Also, Chaudhuri et al.\ reported a loading rate of $2 \times 10^{10}$ atoms/s in a slow, intense, and collimated beam of rubidium atoms \cite{ChaudhuriPRA06} yielding an efficiency of $3.6 \times 10^{11}$ atoms/s/W using 55~mW of cooling laser power. High atom loading efficiency important for laboratory experiments, and even more so for compact and mobile devices, where laser power and space are limited. In this work, we demonstrate a simple 2D atom source where the cold atom loading efficiency in atoms/s/W has been improved by a factor of 1.7 when operating below saturation. We use a 2D~MOT to achieve a loading rate greater than $8 \times 10^{9}$ atoms/s with only 20~mW of laser power;  while the maximum efficiency was $6.2 \times 10^{11}$ atoms/s/W, the highest to our knowledge and a factor of 14 better than our previous multistage 2D source\cite{Ramirez-SerranoOL06}. The atom flux is believed to be much higher than the 3D loading rate. Our atom beam source is compact in physical size, and employs only two retro-reflected elliptical beams aligned normally to the extraction axis. It offers low divergence of the atomic beam without the use of an on-axis push beam and in-vacuum optics~\cite{DieckmannPRA98,BerthoudEurPhys98,JoellenbeckPRA11,ChaudhuriPRA06}, and allows for a much simpler laser system. We currently employ two such atom sources for our gravity gradiometer~\cite{Kohel08} which operate over long data runs of approximately 10 hrs at a time, with un-maintained loading stability of better than 1\% for many months at a time.

The experimental apparatus is illustrated in Fig.~\ref{apparatus}. The high-pressure region is a small ($1.9 \times 1.9 \times 6.4~\text{cm}^{3}$) glass cuvette attached to the ultra-high vacuum (UHV) chamber via a  glass-to-metal seal. The cuvette is completely sealed without any active pumping. A current coil generates the 2D quadrupole magnetic field aligned with the elliptical trapping beams along the extraction axis. A thermoelectric cold finger is attached at the rear of the glass cell in order to maintain Cs vapor pressure for optimum loading. Two fiber coupled collimators forming 12.8$\times$44~mm$^2$ elliptical laser beams are mounted to the coil forms and directed through the vapor cell collection region. Two $\lambda/4$ wave plates with high reflectivity coating on their back side act as retro-reflecting mirrors and are fixed to the coil form parallel to the vapor cell producing the desired $\sigma$-polarizations for the 2D~MOT.

\begin{figure}
\includegraphics[width=0.9\linewidth]{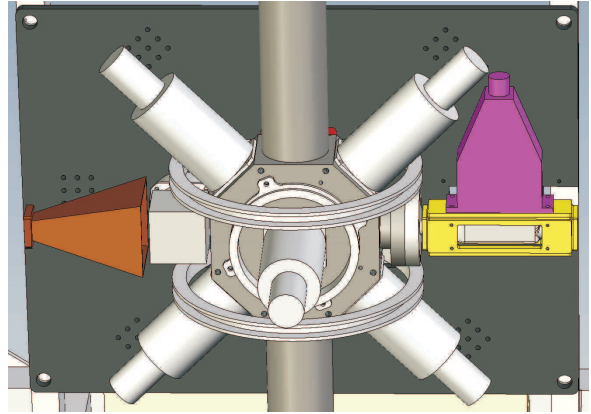}
\caption{CAD drawing of our apparatus with 3D~MOT at left, and 2D~MOT source (without horizontal collimator) on right.}
\label{2Dangle}
\end{figure}

One of the most significant improvements of the present 2D~MOT over conventional elliptical-beam 2D~MOT designs is the angle of the collimators with respect to the axis of the 2D~MOT. Both the horizontal and vertical collimators have been optimized to $4^{\circ}$ tilt as depicted in Fig.~\ref{2Dangle}. Associated retro-mirrors are mounted parallel to the axis of the 2D~MOT, thus insuring that the reflected beams are also projected forward at the same angle as the incident beams, effectively resulting in a pushing component with a fraction of $\sim$14~\% of the overall laser field in each orthogonal direction. The longitudinal field component $k_l$ on one axis is $k_l=2k\sin 4^{\circ}\approx 0.14k$, where $k$ is the wave vector of the incident beam. The factor of 2 arises from the fact that retro-reflection is used. We find no dramatic dependence on the tilting angle over a range of $3$--$15^{\circ}$. The increased push factor at larger angles is offset by reduced lateral trapping efficiency. Nonetheless, a tilt of $4^{\circ}$ showed a slight maximum in flux over this span. It is worth noting that the intrinsic symmetry of a retro-reflected beam setup is very robust and simple to use.

A 4-mm diameter aperture mounted on the end of a 6-cm long differential pumping tube lined with a graphite getter allows the transfer of slow atoms from the 2D~MOT to the UHV trapping region while maintaining a low-conductance path for thermal atoms from the vapor cell source region.  The differential pumping tube extends into the glass cuvette to place the aperture within a few millimeters from the edge of the 2D~MOT
in order to optimize the coupling of the slow atom beam through the differential pumping tube.  The measured background pressure in the UHV region is well below the low pressure limit ($2 \times 10^{-10}$~Torr) for our Bayard-Alpert type ionization gauge.

The entire source apparatus, including the 2D~MOT optics and hairpin coil, is contained within a $23 \times 23 \times 14~\text{cm}^3$ volume.  The required connections include two optical fibers for the trap lasers, and electrical connections for the current coil, a thermoelectric cooler, and RTD temperature monitor.

The 3D~MOT consists of three orthogonal pairs of $\sigma$-polarized laser beams of 24~mm diameter, each with an intensity of 1.7~mW/cm$^2$, aligned with a 3D quadrupole magnetic field in the UHV region.  The center of this trap is located 15~cm downstream from the 2D~MOT.\@ The performance of the 2D~MOT source is characterized by directly measuring the loading rate into the UHV trap.
The capture velocity of the 3D~MOT of 18~m/s is estimated to be lower than the longitudinal mean velocity of the atom beam, thus the observed UHV 3D~MOT loading rates can be interpreted as a lower bound for the actual flux of the atomic beam.
Both MOTs are operated at a fixed detuning of $-1.4\Gamma$, where $\Gamma = 2\pi \times 5.234$~MHz is the natural linewidth of the cesium D$_2$ line. This detuning is found to maximize the steady state atom number in the 3D~MOT.
To simplify the laser system, the same laser frequencies are employed in the 2D and the 3D~MOTs. This limits our ability, however, to to maximize the atom flux by varying the detuning for the 2D MOT source.

\begin{figure}
\includegraphics[width=\linewidth]{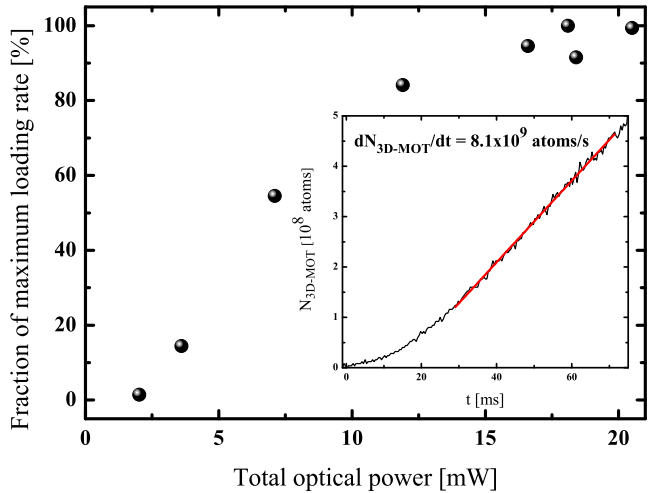}
\caption{Proportional atomic loading rate as a function of total trapping light power in the 2D~MOT. The inset shows the 3D~MOT atom number vs. time, depicting a loading rate greater than 8$\times 10^{9}$~atoms/s.}
\label{intensity}
\end{figure}

Figure~\ref{intensity} shows the atomic loading rate as a function of total optical power in the 2D~MOT and the highest achieved 3D~MOT loading rate, $8.1 \times 10^9$ atoms/s, realized at a magnetic field gradient of 15~G/cm along the radial axes, a detuning of $-1.4\Gamma$ and only 2.3~mW/cm$^2$, or 2.1 times the saturation intensity. Around the maximum intensity available, a saturation is observed in the intensity dependence of the loading rate. Note that because of this saturating behavior our maximum efficiency of $6.2 \times 10^{11}$ atoms/s/W was achieved at a total laser power of only 7~mW. Without tilting the collimators as described above, loading rates were significantly lower than the highest observed with the tilt. In order to rule out that tilting the collimators only assists in aiming through the differential pumping tube, we subsequently moved the current coil form and hence the orientation of the axis of zero magnetic field without any collimator tilt. The highest loading rates could not be reproduced, demonstrating that the effective pushing laser field is the major contribution to the efficiency observed in this work.

We expect that this atomic flux can be further enhanced by increasing the size of the trapping laser beams along with the length of the vapor cell.
An increase in trap length $L$ not only provides an increase in trap volume, but also gives a longer mean cooling time $L/(2 v_{z})$ for atoms with a longitudinal velocity $v_z$. This results in a higher transverse capture velocity and allows atoms with higher longitudinal velocities to be collimated into the atom beam~\cite{SchoserPRA02,Ramirez-SerranoOL06}.
The latter effect results in an increase in the mean velocity for the atom beam, and so this eventually limits further increases in the number of atoms captured from the beam into the UHV MOT~\cite{Ramirez-SerranoOL06}.\@
The total atomic flux, however, continues to increase with the size of the 2D~MOT as long as the mean free path for the slow atoms within the trap region remains longer than the trap length~\cite{SchoserPRA02,Ramirez-SerranoOL06}.
In the current apparatus, the useful atom flux can also be enhanced by decoupling the 2D and 3D~MOT laser frequencies to allow for individually optimized detunings in each trap.

In conclusion, we have demonstrated a high-efficiency compact high-flux 2D~MOT Cs atom source. We showed that higher atom flux may be achieved by angling the trapping collimators with respect to the axis of the magnetic field. This eliminates the need for an additional push beam and also reduces the size and power requirements of the overall package. The measured 3D~MOT loading rate is greater than 8$\times 10^9$ atom/s with a total of 20~mW of trapping laser power, giving a maximum efficiency of $6.2 \times 10^{11}$ atoms/s/W. This performance is highly dependent on other system optimizations. Our total atom flux is believed to be much higher than the 3D~MOT loading rate. The compact design and efficiency of the current apparatus is suitable for cold-atom applications in the laboratory and especially in mobile devices, including cold atom instruments in space.
\\ \\
We thank P. Hamilton for helpful discussions. This research was carried out at the Jet Propulsion Laboratory, California Institute of Technology, under a contract with the National Aeronautics and Space Administration.

\end{document}